# RetraLytix: An Integrated Analytics Dashboard for Mapping Global Trends in Scientific Retractions

Chahat Singh[1], Sejal Gupta[1], Krishna Mundra[1], and Kiran Sharma[1,2*]

[1]*School of Engineering & Technology, BML Munjal University, Gurugram, India*

[2]*Center for Advanced Data & Computational Science, BML Munjal University, Gurugram, India*

*Correspondence*: kiran.sharma@bmu.edu.in*

## Abstract

Retraction is a correction to scientific literature when there is a major flaw, fraud or misuse of ethical practices in the published work. With the increasing growth of research output, number of retracted studies has also increased, which raises concerns about the issue of research ethics and transparency. Moreover, retraction data coming from several platforms or databases limits its scope in tracking the time-to-time retraction trends. To address this, we propose a web-based integrated platform, called *RetraLytix*, for easy analysis of distributed retraction data. It automatically integrates retraction data from major databases like Crossref, Retraction Watch and Open Alex and visualizes data in a user interactive centralized platform. It offers a real-time dashboard, comparative analysis, and benchmarking of entities such as countries, institutions, authors, journals and main research areas. RetraLytix helps users to detect trends, retraction patterns, and assess research environment to make data-driven decisions. The system has a potential to become a research integrity tracking and governance tool for researchers, administrators and policymakers.

## Introduction

Scientific literature serves as the fundamental cornerstone of global research progress, institutional decision-making, and evidence-based policy development. The reliability and integrity of published research are, therefore, critical to maintaining public trust in the academic enterprise. The COPE Retraction Guidelines (2019) provide a standardized framework for retracting published research to maintain integrity in scholarly literature.

While retractions are an essential component of the "self-correcting" nature of science, the dramatic increase in retraction rates across nearly all disciplines in recent decades has raised significant concerns within the scientific community regarding research ecosystems and scholarly accountability. The article by Brainard (2018) states that retractions should not be promoted as punishment instead can be promoted as a transparency and correction in science. It also highlights that the stigma between conscious mistakes like data fabrications, plagiarism, etc. and honest errors should be reported very clearly which will encourage responsible reporting. Steen et al. (2013) highlighted that the rise in retractions are positive and are due to better detection and awareness.

Retraction data is publicly available by Retraction Watch and Crossref. Authors analyzed this data time to time and provide insights about ongoing retractions trends. Sharma (2024) studies the retractions dynamics in India over 2 decades and highlighted the institutional involvement in such retractions. Bar-Ilan et al. (2017) studied that a publication after being retracted still attracts citations. This may be because of the gaps in scholarly communication. After having a sufficiently large literature on studies related to retraction analysis, a holistic view is missing.

Currently, retractions information is usually distributed across multiple publisher websites, hidden within static databases, or distributed via raw metadata files by some database. To compile such information, one need advances in technical skills. Such data fragmentations restrict the holistic picture of retractions to the non-technical stakeholders such as university administrators, journal editors, and policymakers. Existing tools often provide simple metrics but fail to offer interactive, entity-level benchmarking or real-time visualization of evolving trends.

To address this gap, this paper introduces RetraLytix (https://retraction-portal.vercel.app ), an integrated, web-based retraction analytics dashboard designed to enhance transparency in scientific research. RetraLytix transforms raw, fragmented retraction metadata into a centralized, interactive environment. By aggregating data from trusted sources like Crossref database and enriching it via the OpenAlex database, the platform provides users with live dashboards, results comparison tools, and deep-dive insights into the temporal and geographical patterns of retractions.

**Research problem**

The primary challenge in the field of research integrity is not just the increasing number of retractions, but the opacity and fragmentation of retraction data. While datasets like Retraction Watch provide essential records, the information is largely not handy to non-technical stakeholders in a simplified form. Existing web platforms such as PostPub (https://www.postpub.net ), which offers valuable country-wise retraction statistics and basic classifications of retraction reasons but lacks to provide the holistic overview. It lacks to provide a deeper, multi-dimensional analysis at the level of individual authors or journals. Furthermore, existing solutions lack extensive metadata filtering and interactive features that allow users to perform comparisons among research institutions or countries.

To overcome this gap, we are introducing RetraLytix, a unified integrated web tool for retraction data analysis. Table 1 provides a comparative analysis between current services and the proposed RetraLytix platform.

**Table 1.** Comparative analysis between PostPub and RetraLytix.

| Key Features | PostPub | RetraLytix |
|---|---|---|
| Country-wise retraction statistics | ✓ | ✓ |
| Classification of retraction reasons | ✓ | ✓ |
| Author-wise & Journal-wise insights | ✗ | ✓ |
| Side-by-side entity comparison tools | ✗ | ✓ |
| OpenAlex API metadata enrichment | ✗ | ✓ |
| Real-time data exploration | ✓ | ✓ |
| Downloadable datasets or reports | ✗ | ✓ |
| AI chatbot for trend queries | ✗ | ✓ |

**Research Objective**

To develop an integrated retraction analysis web portal, named RetraLytix, that consolidates heterogeneous data sources into a unified dashboard, enabling real-time visualization, comparison among entities, and help in pattern recognition and data-driven decision-making through interactive user interface.

**Methodology**

RetraLytix is developed as a full-stack web application whose architecture ensures the scalability and efficient data flow within the application. The system is mainly structured into three distinct layers: *the frontend layer*, *the backend layer*, and *the data and external services layer* as shown in figure 1. The communication between the user interface and the server is maintained through a secure HTTPS/REST API, which facilitates real-time updates and responsive data rendering across the dashboard.

The frontend layer is supported by React and Vite which provides an interactive environment for data exploration. The user interface is based on Tailwind CSS, and the plots visualization is handled by Recharts and Chart.js. The core of the portal is its backend infrastructure and was implemented using the Flask micro-framework. This layer is responsible for handling API routing, metadata enrichment, and session management.

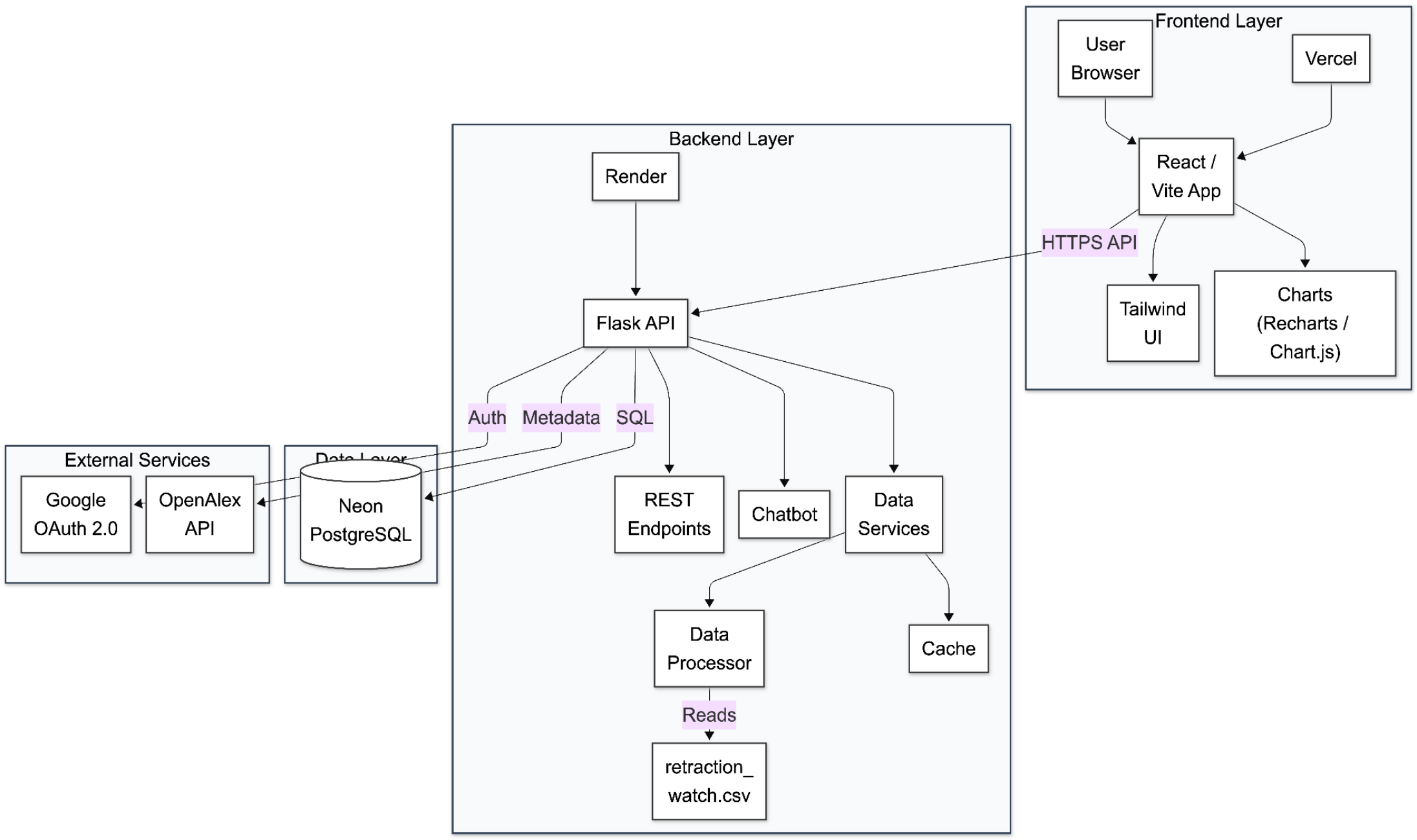

**Figure 1.** Architectural structure of the RetraLytix platform showcasing the interaction between the frontend, backend, and external data layers.

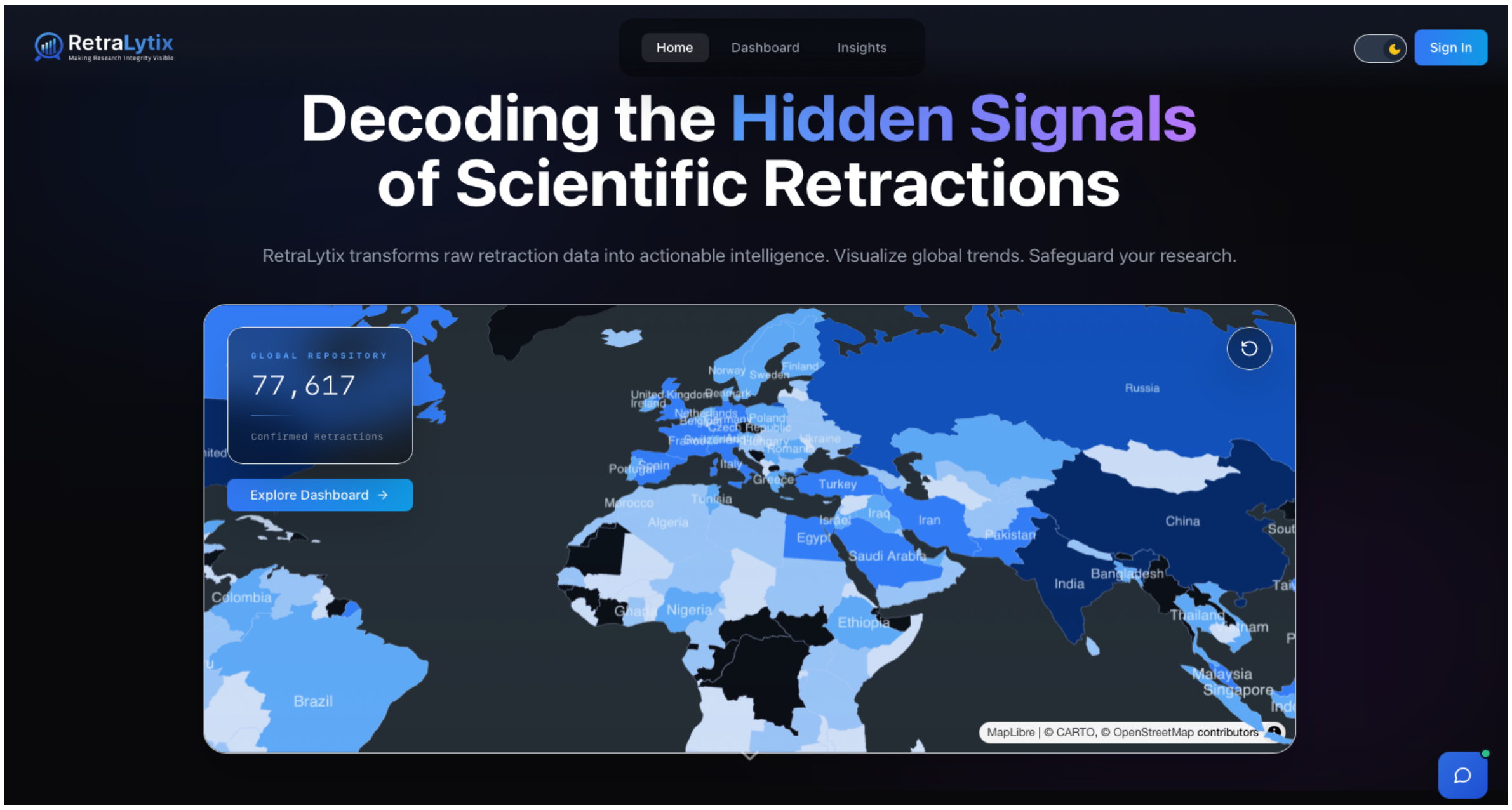

**Figure 2.** RetraLytix portal home page. ([https://retraction-portal.vercel.app](https://retraction-portal.vercel.app))

Further, the system uses the OpenAlex API to fetch supplementary data w.r.t. retraction watch data, such as detailed institutional affiliations and author-level metrics, which are not available in the primary retraction watch dataset. Security and user authentication are managed through Google OAuth 2.0 via the Authlib library, providing a verified login mechanism that protects administrative features. Persistent data, including authenticated user records and system logs, are managed within a Neon PostgreSQL database hosted on AWS infrastructure to ensure the reliability and integrity of the data layer. Figure 2 shows the home page of RetraLytix portal.

## Results and Discussion

Figure 3 displays the RetraLytix comprehensive dashboard which provides a multi-faceted analysis of retraction trends and patterns across the global research landscape. User can analyze this trend over different factors like reasons, subjects, journals, publishers, etc. A "Time to retraction" component highlights the number of retractions over the duration of 0-6 months., 6-12 months, 1-2 years, and so on. A dedicated summary section presents key statistics and trends such as average, median and means time of retraction, empowering researchers and administrators to identify at-risk institutions and emerging patterns in research integrity.

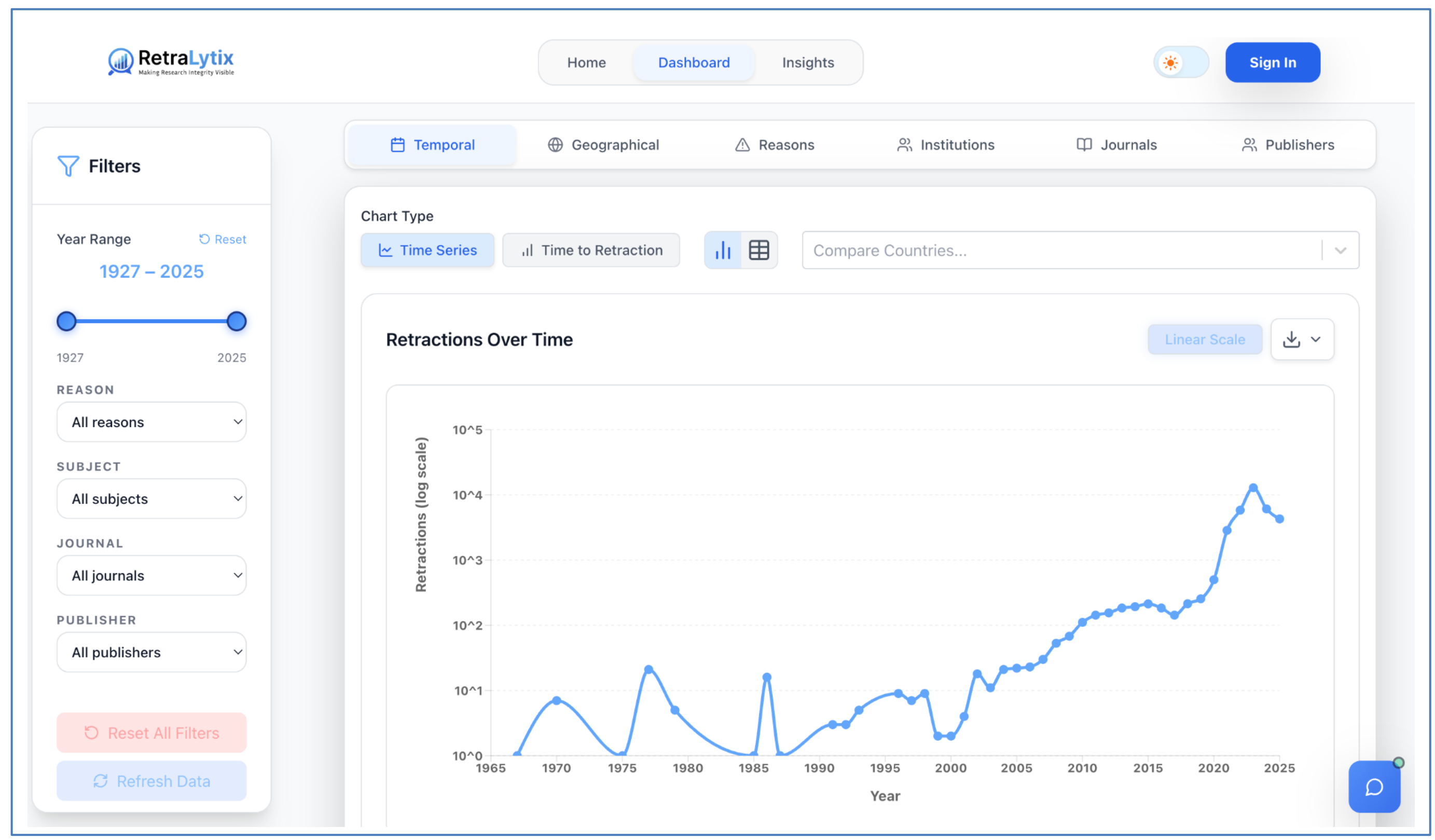

**Figure 3.** Overview of dashboard. The dashboard comprises of temporal, geographical, retraction reasons, institutions, journals, and publishers' analysis.

Figure 4 (left) shows the comparative analysis of retraction growth across countries. China shows upward trend as compared to India and USA. This is to note that this data is not normalized at a moment as per number of actual publications. This display uses a raw retraction count. Similarly, Figure 4 (left) shows the distribution of various retraction reasons categories.

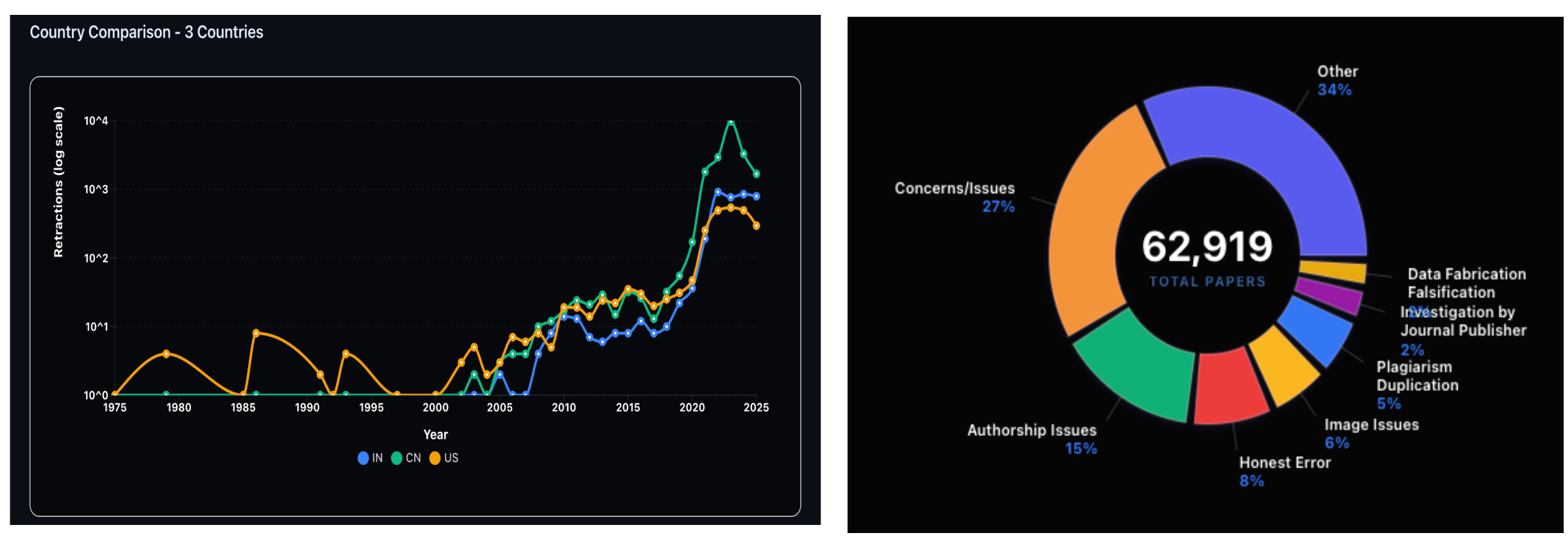

**Figure 4.** (*left*)Temporal analysis of retracted publications among India, China and USA. (*right)* Retraction reasons distribution.

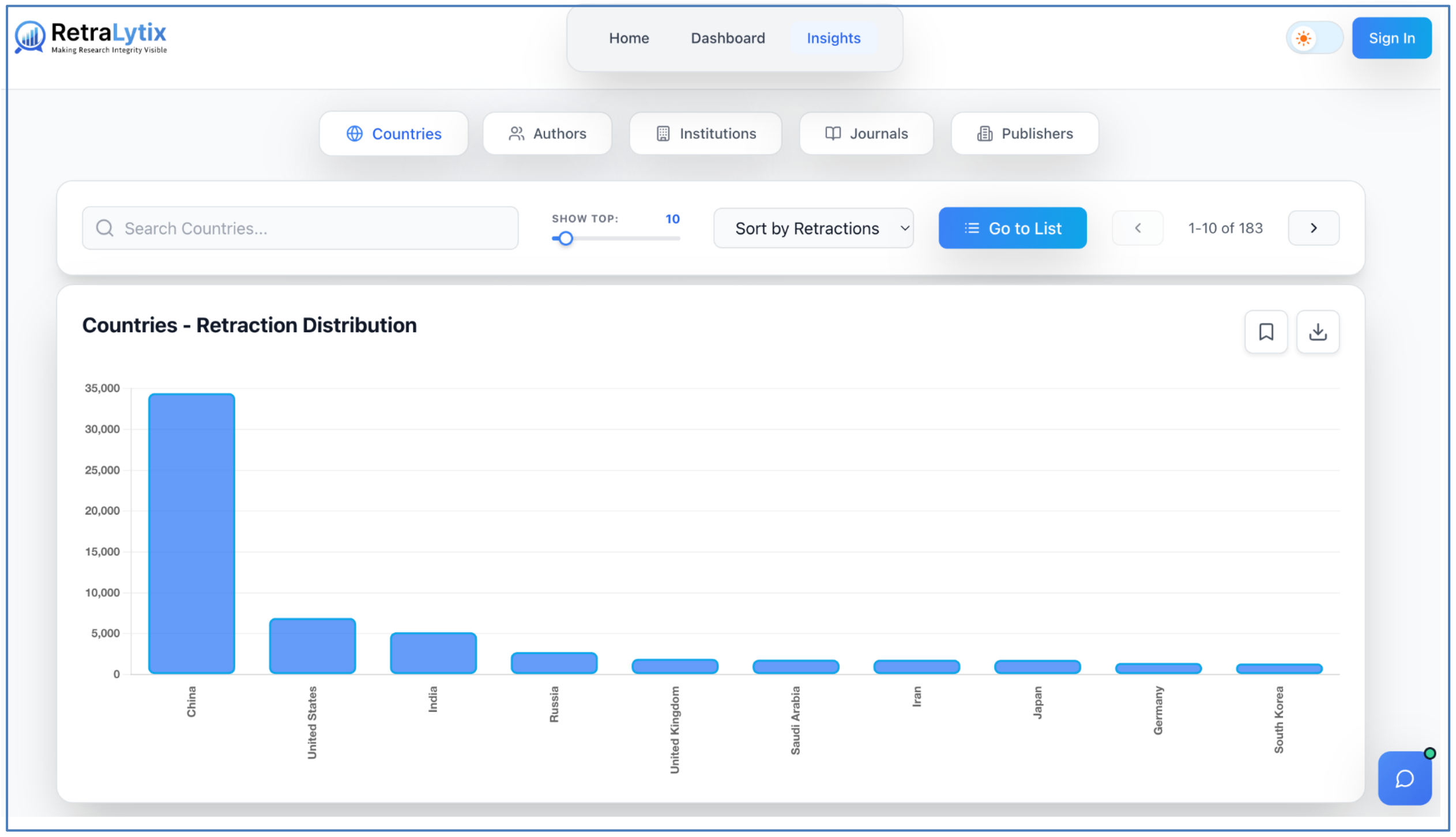

**Figure 5.** Demonstration of RetraLytix insights page. The insights page comprises of entities like countries, author, institutions, journals, and publishers.

The Insights page shown in figure 5 transforms raw retraction data into actionable intelligence through advanced analytical visualizations. This page highlights the thematic patterns in research misconduct, track institutional reputation trajectories over custom time periods, and conduct comprehensive comparative analysis between peer institutions and funding bodies. The page features interactive plots and trend analysis that reveal geographic and temporal hotspots of retractions. The retraction reasons breakdown enables stakeholders to understand the underlying causes within that country, journal, institution, etc. Further the analyzed plot customization allows researchers and administrators to generate evidence-based reports for further insights.

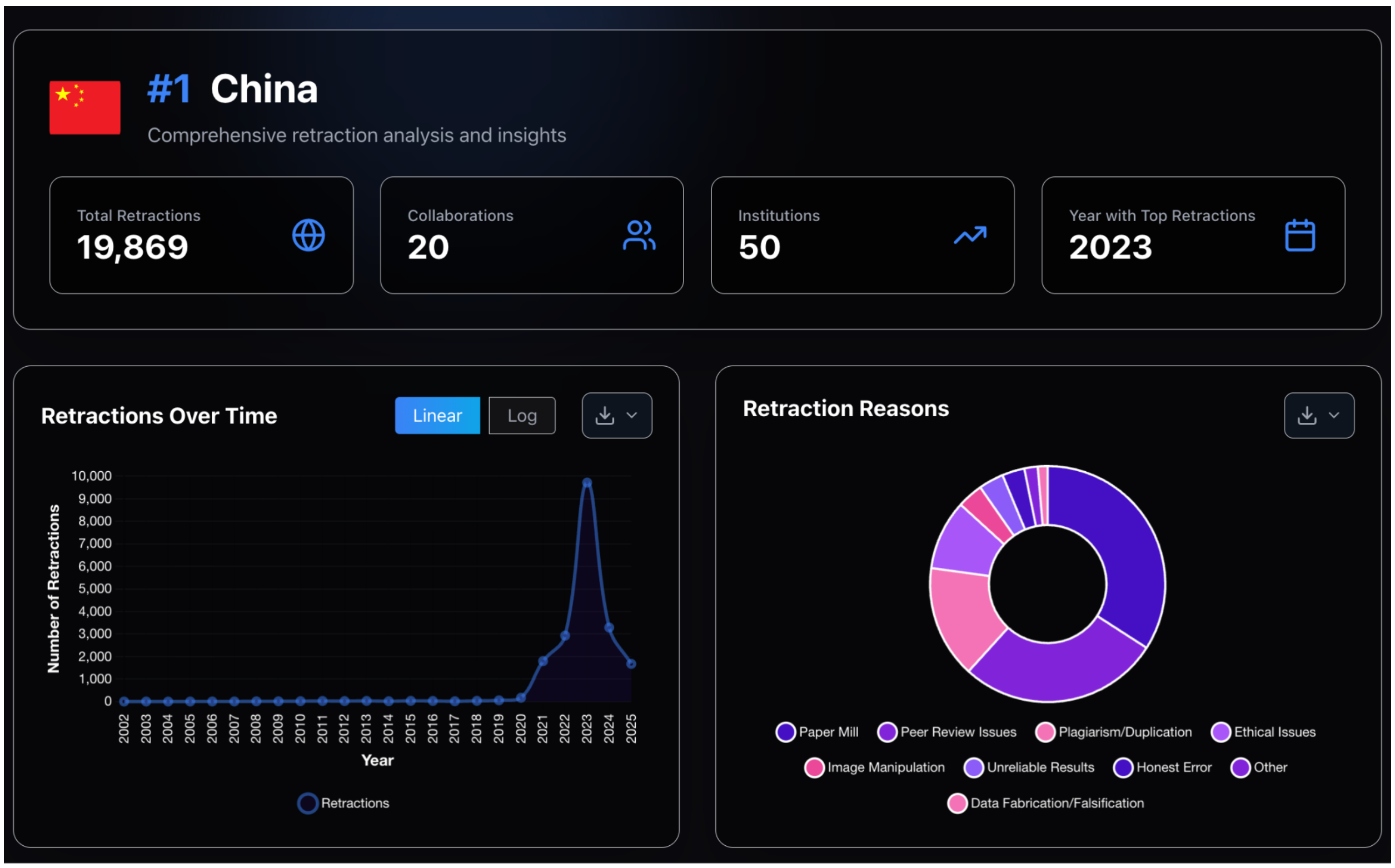

**Figure 6.** Glimpse of RetraLytix insights of a country - China.

Figure 6 displays the retraction information of top listed country China with high number of reactions. This page provides a complete picture of retraction trends, reasons of retraction, institute involve from China, their domestic and international collaboration, their discipline wise retraction records, and so on. Further, one can dig deeper to investigate the authors under these retraction categories from that country and from which journal and under which publishers all their retractions are reported. Such detailed and deep information on specific country provides a holistic view of retractions performed in that country.

## Conclusion

RetraLytix is a customized solution to the problems of unintelligible information of retractions that contributes to a holistic, interactive and real-time analytical platform for tracking research integrity globally. Such an integrated view is needed in multiple ways like live visualization of the growth, to dig down deep to have an insight into the reason, to find out which country or institute is more involved, or which journal needs to be very careful with their research and publishing policy to reduce retractions. At present, none of these phenomena are available with a holistic view. RetraLytix is a user interactive platform which delivers this unified view. RetraLytix can be employed as a policy level support system for policymakers/administrative bodies to make informed decisions.

## Future Scope

The future development of RetraLytix will focus on (i) the integration of Large Language Models (LLMs) into chatbot to provide users with analysis; (ii) to implement bookmarking feature which will allow users to save and export specific graphs and compile a report; (iii) to integrate a reference-scanning tool to identify "zombie citations" before publication; (iv) to perform field normalization to ensure fair benchmarking across disciplines; (v) to classify all retractions based on SDGs to understand which category dominates; (vi) to study the dynamics of citations received by retracted papers.

**Conflict of Interest**
The authors declare no conflict of interest.

## Data Availability

The data is solely taken from Crossref (https://api.crossref.org/), Retraction Watch (https://retractionwatch.com/) annd mapped with OpenAlex API (openalex.org).

## Authors Contribution

1. Dr. Kiran Sharma: Conceptualization, methodology, and supervision as Principal Investigator.
2. Chahat Singh: Frontend development and system architecture design.
3. Krishna Mundra: Backend infrastructure and API development/integration.
4. Sejal Gupta: Implementation of Google Auth security protocols and full-stack development support.

## References

1. Brainard, J. (2018). Rethinking retractions. *Science*, 362(6413), 390–393.

2. Retraction Watch. (2023). *Retraction Watch Database*. Center for Scientific Integrity. Available at: https://retractionwatch.com/
3. Crossref. (2023). *Crossref Metadata API*. Available at: https://api.crossref.org/
4. OpenAlex. (2023). *OpenAlex API Documentation*. Available at: https://openalex.org/
5. PostPub. (2023). *Post-publication peer review and retraction statistics platform*. Available at: https://www.postpub.net/
6. Sharma, K. (2024). Over two decades of scientific misconduct in India: Retraction reasons and journal quality among inter-country and intra-country institutional collaboration. *Scientometrics*, *129*(12), 7735-7757.
7. Steen, R. G., Casadevall, A., & Fang, F. C. (2013). Why has the number of scientific retractions increased? *PLoS ONE*, 8(7), e68397.
8. Bar-Ilan, J., & Halevi, G. (2017). Post retraction citations in context: A case study. *Scientometrics*, 113(1), 547–565.
9. COPE (Committee on Publication Ethics). (2019). *Retraction Guidelines*. Available at: https://publicationethics.org/
10. Google. (2023). *OAuth 2.0 Authorization Framework*. Available at: https://developers.google.com/identity/protocols/oauth2